\documentclass[twocolumn]{svjour3}
\usepackage[utf8]{inputenc}
\usepackage{graphicx}
\usepackage{amsmath}
\usepackage{color}
\usepackage{enumerate}
\usepackage{todonotes}

\usepackage{url}
\usepackage{hyperref}

\begin{document}

\title{Generalization of Higuchi's fractal dimension for multifractal analysis of time series with limited length%\thanks{Grants or other notes
}


\author{Carlos Carrizales-Velazquez         \and
        Reik V. Donner \and Lev Guzm\'an-Vargas  %etc.
}

\institute{C. Carrizales-Velazquez \and L. Guzm\'an-Vargas \at
              Unidad Profesional Interdisciplinaria en Ingenier\'ia y Tecnolog\'ia Avanzada, Instituto Polit\'ecnico Nacional, Ciudad de M\'exico, M\'exico. \\
              %\email{c.carrizales.v@gmail.com} \\
              \email{lguzmanv@ipn.mx}
                        %  \\
           \and
            Reik V. Donner \at
              Department of Water, Environment, Construction and Safety, Magdeburg-Stendal University of Applied Sciences, Magdeburg, Germany, and \\
              Research Department I -- Earth System Analysis \& Research Department IV -- Complexity Science, Potsdam Institute of Climate Impact Research (PIK) -- Member of the Leibniz Association, Potsdam, Germany. \\
              %\email{redonner@pik-potsdam.de}
           %\and
           	%T. Author \at
           	  %Unidad Profesional Interdisciplinaria en Ingenier\'ia y %Tecnolog\'ia Avanzada, Instituto Polit\'ecnico Nacional, Ciudad de M\'exico, M\'exico. 
           	  %\email{lguzmanv@ipn.mx}
%           	  \emph{Present address:} of T. Author
           	  %Tel.: +52-55-57296000 extensi\'on 56873\\
}

\date{Received: date / Accepted: date}

\maketitle

\begin{abstract}
We introduce a generalization of Higuchi's estimator of the fractal dimension as a new way to characterize the multifractal spectrum of univariate time series. The resulting multifractal Higuchi dimension analysis (MF-HDA) method considers the order-$q$ moments of the partition function provided by the length of the time series graph at different levels of subsampling. The results obtained for different types of stochastic processes as well as real-world examples of word length series from fictional texts demonstrate that MF-HDA provides a reliable estimate of the multifractal spectrum already for moderate time series lengths. Practical advantages as well as disadvantages of the new approach as compared to other state-of-the-art methods of multifractal analysis are discussed, highlighting the particular potentials of MF-HDA to distinguish mono- from multifractal dynamics based on relatively short time series.
\end{abstract}

\section{Introduction}
\label{intro}
Since Benoit Mandelbrot applied the concept of fractality for the first time to real-world time series, different approaches have been proposed to detect fractal and multifractal scaling properties in diverse signals from complex systems \cite{mandelbrot1983fractal,janssen1998statistics,feder2013fractals,frisch1985singularity,Stanley1988,Ivanov1999} (for recent reviews, see~\cite{Schmitt2016,Jiang2019}). In many real-world cases, this scaling behavior reflects the fact that embedded variability components cover a wide range of time scales despite the absence of a dominant periodicity \cite{Kantelhardt2011,bunde2012science}. For monofractal time series, a single scaling exponent is sufficient to completely characterize the fractal properties of the dynamics displayed by the analyzed sequence. By contrast, for multifractal signals a variety of scaling exponents are necessary to fully describe the dynamics. 

The multifractal formalism was originally introduced in the context of turbulence studies \cite{frisch1985singularity,kolmogorov1941local,frisch1995turbulence,mccauley1990introduction,coleman1992introduction} and velocity fluctuations \cite{frisch1985singularity}. Later, Mandelbrot \cite{mandelbrot1983fractal,mandelbrot1989multifractal} described the multifractal properties of geometric objects, where the diversity of local scaling exponents characterizes the multifractality (broad spectrum of exponents). Nowadays, multifractal analysis has became a universal technique for data analysis, with a great variety of applications across a multitude of fields \cite{janssen1998statistics,mccauley1990introduction,coleman1992introduction,mandelbrot1999multifractal,paladin1987anomalous,olemskoi2000theory,touchette2009large,kwapien2012physical}. 

In the context of complex time series, the two standard methods to detect multifractality are based on either the construction of a partition function or the concept of extended self-similarity associated with the notion of structure functions (for details, see~\cite{feder2013fractals,Jiang2019}). Both approaches provide direct procedures to obtain the spectrum of multifractal scaling exponents involved in the multifractal measure. While the partition function method is mainly rooted in the box-counting approach \cite{feder2013fractals} and may therefore require relatively long time series for proper evaluation, structure functions rely on the nonlinear scaling behaviors of increments bridging increasingly large time differences \cite{Anselmet1984,barabasi1991multifractality}. A more sophisticated method for multifractal analysis, the wavelet transform modulus maxima (WTMM) method, has been introduced by Arneodo \emph{et~al.}~\cite{MMWT,muzy1994multifractal} and applied to a great variety of problems from various disciplines. Here, the wavelet transform is used to obtain the sequence of local maximum values over a range of scales, which also accounts for the wide-spread nonstationarity of real-world time series. In a similar spirit, a conceptually related approach based on time scale decomposition by means of a more data-adaptive technique (the empirical mode decomposition) has been developed more recently \cite{Welter2013,Alberti2019}.

Another state of the art approach of multifractal analysis has been introduced by Kantelhardt \emph{et~al.}, who proposed a generalization of detrended fluctuation analysis \cite{MFDFA} to obtain an estimate of the multifractal spectrum of nonstationary time series. This multifractal detrended fluctuation analysis (MF-DFA), together with various algorithmic variants thereof differing in the way of how local time series detrending is performed, provides a robust way to determine the multifractal scaling characteristics in terms of generalized Hurst exponents $h(q)$, where $q$ represents the order of the statistical moment under study. However, both WTMM and MF-DFA exhibit certain practical challenges in determining the corresponding multifractal spectra from relatively short time series \cite{Gieraltowski2012,Waveletvsmdfa2006}, which may lead to spurious identification of multifractality. 

As a possible alternative, this paper introduces a generalization of Higuchi's fractal dimension (HFD) estimator for univariate time series \cite{higuchi1988approach,Higuchi1990}, which has been widely used as a technique for quantifying mono\-fractal scaling characteristics in nonstationary time series from different areas of science, ranging from geophysics \cite{higuchi1988approach,Nikolopoulos2020,Ramirez2008,Donner2015,Cuomo1999} to biomedicine \cite{Kesic2016,Guzman2003,Schmitt2007,Contreras2017}. Specifically, we introduce the concept of multifractal Higuchi dimension analysis (MF-HDA), which is applicable to comparatively short sequences and provides very stable estimates of scaling exponents for the construction of the multifractal spectrum. 

The remainder of this paper is organized as follows. In Section 2, we briefly review some essential details on the HFD method and introduce a thorough modification of the original approach to obtain the corresponding scaling exponent. Section 3 describes the multifractal generalization of the monofractal dimension estimator. Its application to simulated time series from two illustrative stochastic model systems and a subsequent comparison of the obtained performance with that of MF-DFA are presented in Sections 4 and 5. In Section 6, we then apply our method to some real-world examples of world length sequences in English fictional texts. Finally, a discussion and some conclusions and general remarks are provided in Sections 7 and 8, respectively.

\section{The fractal dimension of irregular time series}
\label{sec:1}
\subsection{Higuchi's estimator of the fractal dimension}
\label{sec:2}
Consider a time-series $X(t)$ with $t=1,2,\ldots,N$. First, we construct a set of sub-sampled time series 
\begin{equation}
\begin{split}
    X_k^m=\Big(X(m),X(m+k),&X(m+2k),\ldots,\\
    &X\Big(m+\Big\lfloor\frac{N-m}{k}\Big\rfloor\cdot k\Big)\Big) 
    \end{split}
\end{equation}
\noindent
at different scales (determined by $k$) with $m=1,2,\ldots,k$, where $\lfloor\bullet\rfloor$ denotes the (lower) integer part of a real number. Here, $k$ and $m$ are integers which describe a time lag and the initial time index, respectively. Next, the length of the curve associated with a given sequence $X_k^m$ is defined as \cite{higuchi1988approach}
\begin{equation}
    L_m(k)=\Bigg\{ \Bigg( \sum_{j=1}^{j_{max}} \Delta X_{k}^{m}(j)  \Bigg) \frac{N-1}{j_{max} \cdot k} \Bigg\}\Bigg/ k,
    \label{L-Hig}
\end{equation}
where $\Delta X_{k}^{m}(j)=|X(m+jk)-X(m+(j-1)\cdot k)|$ are the absolute increments of the subsampled series, $j_{max}=\left\lfloor\frac{N-m}{k}\right\rfloor$ is the total number of increments at scale $k$, and $(N-1)/(j_{max}\cdot k)$ represents a normalization factor. 

We notice that the quantity $\Delta X_{k}^{m}$ represents the differences corresponding to each $k$-lag value, which when aggregated and normalized by $j_{max}$ %$(N-m)/k$ 
gives the average of the increments for that scale $k$. For the purpose of the present study, consider for example $X(t)$ to represent the position of a one-dimensional random walk at time $t$. Then, $\Delta X_k^m$ is the absolute displacement within a time step $k$ with $m=1,2,\ldots,k$, where the best possible temporal resolution corresponds to $k=1$, the resolution decreases by one half for $k=2$, and so on. Thus, the mean absolute displacement of $X(t)$ starting at $m$ at a resolution $k$ is given by 
\begin{equation}
\langle \Delta X_k^m \rangle =\frac{1}{j_{max}} \sum_{j=1}^{j_{max}} \Delta X_{k}^{m}(j) ,
\label{inc}
\end{equation}
and the total mean displacement is calculated as $\ell_m(k)=\frac{N-1}{k} \langle \Delta X_k^m \rangle$, where $\frac{N-1}{k}$ represents the typical number of displacements available at a resolution $k$. In order to make these total mean displacements obtained for different resolutions $k$ mutually comparable, the total length of the curve describing the time series graph coarse-grained at scale $k$ and starting at $m$ is defined as $L_m(k)=\ell_m(k)/k$, and the total mean curve length at scale $k$ is given as $L(k)=\langle L_m(k)\rangle$ \cite{higuchi1988approach,Higuchi1990}.

In case of a self-affine (i.e., fractal) time series, it has been demonstrated that
\begin{equation}
    L(k) \sim k^{-D_A},   
\end{equation}
\noindent 
where $D_A\in(1,2)$ is an estimate of the fractal dimension of the original time series graph and the subscript $A$ stands for averaging \cite{higuchi1988approach,Higuchi1990}.

\subsection{Modified HFD estimator}
\label{Exp-Hig-Length}

For the purpose of generalizing the HFD method to a multifractal formalism, it is necessary to first introduce an alternative formulation of an estimator of the fractal dimension that is based on the same rationale as Higuchi's original method. Here, our modified approach replaces the averaging procedure for calculating the mean absolute displacement by an expectation value in terms of probabilities of increment sizes $\Delta X_k^m$ (Eq.~\ref{inc}). 

We first recall that the number of increments $\Delta X_{k}^m$ at scale $k$ is $\lfloor(N-m)/k\rfloor$, and the total number is $\sum_{m=1}^{k}\lfloor (N-m)/k\rfloor$, which equals $N-k$ \cite{graham1989concrete}. For numerically evaluating the corresponding expectation values, we follow here a simple histogram based approach (other more sophisticated density estimators may be used as well, yet we focus here on this simple strategy and outline corresponding follow-up studies on this aspect as parts of future work). Let us consider $P_n(\bullet)$ representing the discrete probabilities of increments following into disjoint intervals $\Delta X_n(k)$, where $n$ represents the $n-$th interval of an $N_b$ equiprobable partition of the entire range of increment values in increasing order, that is, each interval (bin) contains approximately a fraction of $1/N_b$ of the $N-k$ increments, i.e., $P_n(\Delta X)\approx 1/N_b$. Note that in practice, the total number of increments $\Delta X$ may not be an integer multiple of $N_b$, which results in minor deviations among class frequencies that become gradually less important as $N_b/N\to\infty$. For the purpose of the present work, we employ numerical estimates of the respective $n/N_b$ quantiles ($n=1,\ldots,N_b-1$) to define mutually disjoint intervals with approximately the same population.

In order to obtain a stable estimate of the expectation value of the increments, we start with calculating the ``local-mean'' increment values, i.e., the mean values for each bin, as $\langle \Delta X_n(k)\rangle =  \frac{1}{f_n}\sum_{s=1}^{f_n} \Delta X_s(k)$, where $f_n$ represents the actual number of increments within the $n$-th bin and $\sum_n f_n=N-k$. Then, the expectation value of the increments at scale $k$ can be estimated as
\begin{equation}
    E[\Delta X(k)] = \sum_{n=1}^{N_b}  \langle \Delta X_n(k)\rangle  P_n(\Delta X(k)).
    \label{S_expectation}
\end{equation}
\noindent 
In analogy with Higuchi's original method, the total length of the curve at resolution $k$  can then be written as: 
\begin{equation}
    \mathcal{L}(k) = \frac{N-1}{k^2} E[\Delta X(k)].
\end{equation}
\noindent
Again, we may expect that if the original time series is fractal, then the total length follows a power-law behavior $\mathcal{L}(k) \sim k^{-D_{E}}$, where $D_{E}$ is our modified Higuchi type estimator of the fractal dimension and the subscript $E$ stands for expectation.

\begin{figure}[t]
    \includegraphics[width=0.5\textwidth]{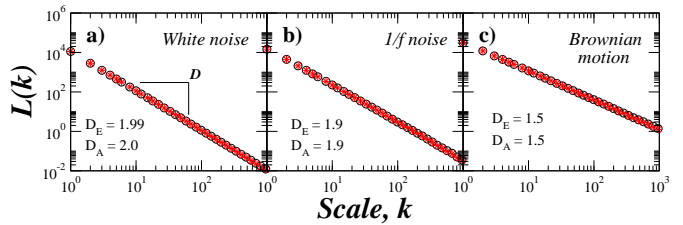}
    \caption{Estimation of Higuchi's fractal dimension for three different types of elementary stochastic processes by using averaging (original method) and numerical approximations of the expectation values (modified approach) of the respective curve lengths. a) White noise (spectral exponent $\beta = 0$), b) $1/f$ noise ($\beta=1$), and c) Brownian motion ($\beta = 2$). The fractal dimension estimate obtained by the expected curve length is denoted as $D_E$, and that obtained using Higuchi's original method (average length) as $D_A$. The obtained numbers correspond to averages taken over 50 independent realizations of length $N=25,000$.}
    \label{EA_Displa}
\end{figure}

In order to verify the consistency of our modified approach with the classical Higuchi estimator, we applied both our methodology and the original HFD method to some elementary stochastic time series with known properties. Specifically, we first consider realizations of three representative monofractal processes: uncorrelated noise (white noise), intermittent long-term correlated noise ($1/f$ noise) and classical Brownian motion \cite{mandelbrot2002gaussian}. The respective results are summarized in Fig.~\ref{EA_Displa}. In general, we observe an excellent agreement between our proposed modification and Higuchi's original method for all three monofractal series.

\begin{figure}[h]
    \includegraphics[width=0.5\textwidth]{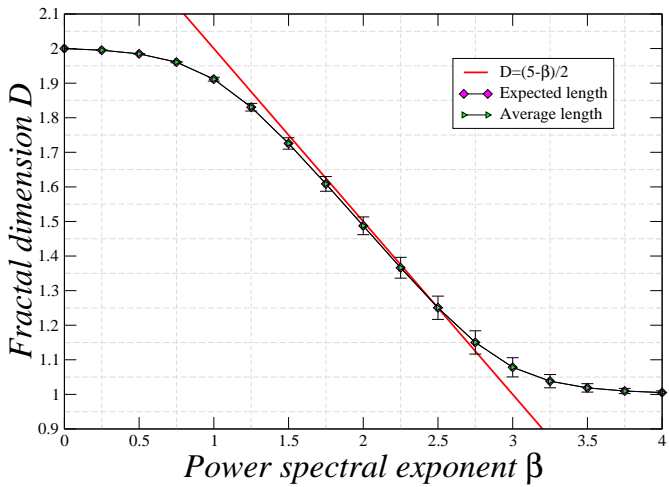}
    \caption{Estimates of the fractal dimension ($D_E$ and $D_A$) for monofractal series of fractional Gaussian noises with prescribed spectral exponent $\beta$. Both methodologies lead to very similar average values for the simulated stochastic processes. The dashed line indicates the theoretical relationship between the fractal dimension and the spectral exponent $\beta$ for $N\to\infty$, which is given by $D=(5-\beta)/2$ and applies for $1<\beta<3$ \cite{Higuchi1990,Guzman2005}. Error bars represent the numerical standard deviations obtained from 100 independent realizations of length $N=10,000$.}
    \label{set_mono}
\end{figure}

To further evaluate the agreement between both methodologies, we systematically generated monofractal time series of fractional Gaussian noise with gradually varying spectral scaling exponent $\beta$ \cite{rangarajan2000integrated} and then applied both approaches for estimating the corresponding fractal dimensions. Figure~\ref{set_mono} shows the numerical results for different processes with $\beta\in (0,4)$. Note that we have to expect appropriate estimates to exist only for $\beta\in (1,3)$ and the finite-sample estimates to become gradually biased as any of the boundaries of the latter interval are approached \cite{Witt2013}. This expectation is met by our numerical results, which again demonstrate that both methodologies lead to very similar fractal dimension estimates.

\section{Multifractal Higuchi Dimension Analysis}\label{Sec-MDA}

\subsection{Generalization of the modified Higuchi estimator}

Based on our modification of Higuchi's fractal dimension estimator as presented in the previous section, our principal interest is the generalization of this monofractal analysis method to a multifractal framework. For this purpose, we consider the $q$-th order moment of the modified Higuchi's curve length as
\begin{equation}
    \mathcal{L}(k,q) = \frac{N-1}{k^2} \Bigg\{ \sum_{n=1}^{N_b} \Big[\frac{1}{f_n}\sum_{s=1}^{f_n} (\Delta X_s(k))^q \Big]  P_n(\Delta X(k)) \Bigg\}^{1/q}.
    \label{Lkq}
\end{equation}
\noindent 
Here, $q$ can take any real number (for a discussion of the special case of $q=0$, see Appendix A). The behavior of this generalized Higuchi length $\mathcal{L}(k,q)$ for different orders of the moment $q$ is indicative of a dependence of the characteristic (fractal) scaling exponents on $q$. More precisely, if the time series under study displays power law correlations (i.e., long-term memory resulting in fractal scaling characteristics), we can expect to find the power law scaling behavior
\begin{equation}
    \mathcal{L}(k,q) \sim k^{-d_q},
    \label{Scal}
\end{equation}
\noindent 
where $d_q$ is a generalized fractal dimension associated to the $q$-th moment. Notice that for $q=1$, our modification of the standard Higuchi fractal dimension method is recovered. For Gaussian time series exhibiting an extended self-similarity (multifractal) property, $d_q$ is related to the generalized Hurst exponent $h_q$ (also known as the H\"older exponent) by the relationship $h_q=2-d_q$ \cite{bunde2012science}. In our case, the singularities are identified by the increments $\Delta X_s(k)^q$. For $q>0$, the contributions of fluctuations corresponding to large increments are highlighted, while the opposite situation is obtained for $q<0$, i.e., fluctuations associated to small increments are amplified.

We emphasize that the increments $\Delta X_s(k)$ can be small enough to affect the numerical estimates of the moments of order $q$, especially for negative $q$, where reliable estimation becomes more and more difficult. To address this challenge, a suitable numerical regularization procedure may become necessary to allow a feasible calculation of the generalized lengths in Eq.~\eqref{Lkq}. If the local behavior of the generalized length $\mathcal{L}(k,q)$ is found to be unstable (see Appendix B for details), we suggest here to replace the local mean $\langle \Delta X_n(k)\rangle=  \frac{1}{f_n}\sum_{s=1}^{f_n} \Delta X_s(k)$ in Eq.~\eqref{Lkq} by a regularized version trimmed at small increment values, $\langle \Delta X_{n'}(k)\rangle _{>p_r}$, where $n'$ represents the $n'-th$ interval of a new equiprobable partition in which we have removed all increment values below their empirical $r-$th percentile $p_r$. In this case, the number of increments at scale $k$ is now given by $\sum_{n'} f_{n'}=N-k-N_{p_r}=(1-p_r/100)(N-k)$ with $N_{p_r}=p_r(N-k)/100$ denoting the number of elements below the $r-$th percentile. In our numerical experiments described in the remainder of this work, we typically set a percentile value within the range $r\in [1,15]$ (see Appendix B), while for the number of bins in the partition, we set $N_b=16$ (where for very long time series, finer partitions may also become numerically feasible).

\subsection{Partition function and multifractal analysis}

Considering the scaling behavior suggested in Eqs.~\eqref{Lkq} and \eqref{Scal}, 
\begin{equation}
     \frac{N-1}{k^2} \Bigg\{ \sum_{n=1}^{N_b} \Big[\frac{1}{f_n}\sum_{s=1}^{f_n} (\Delta X_s(k))^q \Big]  P_n(\Delta X(k)) \Bigg\}^{1/q} \sim k^{-d_q},
\end{equation}
\noindent
it follows that
\begin{equation}
    \sum_{n=1}^{N_b} \left[\frac{1}{f_n}\sum_{s=1}^{f_n} (\Delta X_s(k))^q \right] P_n(\Delta X(k))  \sim k^{q (2-d_q)}.
    \label{Z1}
\end{equation}
\noindent
Using the fact that $P_n(\Delta X(k)) = f_n / (N-k)$ and that for self-affine time series, $h_q=2-d_q$, Eq.~(\ref{Z1}) can be rewritten as
\begin{equation}
    \frac{1}{N-k} \sum_{i=1}^{N-k} (\Delta X_i(k))^q \sim k^{q h_q},
    \label{Z2}
\end{equation}
\noindent 
which represents the well-known generalized structure function \cite{feder2013fractals}. We notice that asymptotically (i.e., for large $N$) $1/(N-k) \sim k$, and therefore, 
\begin{equation}
    \sum_{i=1}^{N-k} \Delta (X_i(k))^q \sim k^{q h_q-1}.
    \label{Z4}
\end{equation}

In the standard multifractal formalism, the partition function $Z(q,k)$ is used to define the mass exponent $\tau(q)$, and the following scaling relation is observed:
\begin{equation}
    Z(q,k)  \equiv \sum_{i=1}^{N-k} (\Delta X_i(k))^q \sim k^{\tau(q)},
    \label{Part}
\end{equation}
\noindent
where we have assumed that the left-hand of Eq.~(\ref{Z4}) corresponds to the partition function $Z(q,k)$. By comparing Eqs.~ (\ref{Z4}) and (\ref{Part}), we obtain the relationship between the scaling exponents as 
\begin{equation}
 \tau(q) = qh_q-1. 
 \label{mass}
 \end{equation}
\noindent
Thus, we have related the exponent $h_q$ to the fractal scaling exponent $\tau(q)$, and the requirement of the fractal dimension value (when $q=0$) for the geometric support of the multifractal measure $-\tau(0)=1=D_f$ is fully satisfied \cite{MFDFA}. Moreover, the canonical way to characterize a multifractal signal is the singularity spectrum $f(\alpha)$, which is obtained by means of $\tau(q)$ and its first derivative $\alpha$ as 
\begin{equation}
\alpha = \tau'(q)\hspace{0.4cm} \quad \Rightarrow \quad \hspace{0.4cm} f(\alpha)=q\alpha-\tau(q),
\end{equation}
\noindent
where $f(\alpha)$ represents the fractal dimension of the subset of the signal characterized by the local exponent $\alpha$ \cite{feder2013fractals}.

\section{Numerical examples: stochastic time series}

\subsection{Monofractal fractional Gaussian noises}
\begin{figure}[h]
    \includegraphics[width=0.5\textwidth]{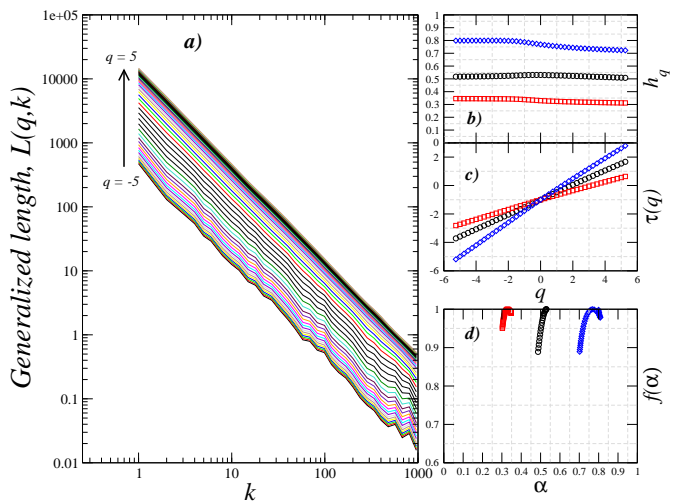}
    \caption{Multifractal Higuchi dimension analysis of monofractal fractional Gaussian noise time series of length $N=10,000$ with $H=0.3$, $H=0.5$ and $H=0.75$. a) Generalized curve lengths $\mathcal{L}(k,q)$ as functions of the scale $k$ for the case $H=0.5$ for several $q$-values within the interval $-5 \leq q \leq 5$. b) Scaling exponents $h_q$ as a function of $q$, obtained from estimates of the slopes ($d_q$) of the curves in a), for scales $1 \leq k \leq 10^3$. The results show only a weak dependence on $q$, confirming the monofractality of the series. c) Behavior of $\tau(q)$ versus $q$, which indicates a linear dependence. d) Multifractal spectra $f(\alpha)$. Here, the singularity spectra $f(\alpha)$ were obtained from the Legendre transform of $\tau(q)$.}
    \label{BN_Res}
\end{figure}

In order to demonstrate the performance of our multifractal Higuchi dimension analysis (MF-HDA) method when applied to simulated monofractal times series, we generated realizations of fractional Gaussian noises with prescribed Hurst exponents of $H=0.3$, $H=0.5$ and $H=0.75$, respectively, by means of the Fourier transform method \cite{rangarajan2000integrated}.  Figure~\ref{BN_Res} shows the results of the multifractal analysis for all three processes. As it can be seen in Fig.~\ref{BN_Res}a, the generalized curve lengths for $H=0.5$ exhibit a power law scaling behavior for which well defined exponents can be estimated (Eq.~\ref{Scal}). In Fig.~\ref{BN_Res}b, the exponents $h_q=2-d_q$ are plotted against the moment $q$. As expected, only a weak dependence of the numerical estimates on $q$ is observed, indicating that the time series are, in fact, monofractal. Moreover, Fig.~\ref{BN_Res}c shows the behaviors of $\tau(q)$ versus $q$, where clearly linear dependencies on $q$ are observed for all three processes. These results lead to very narrow multifractal spectra estimates (Fig.~\ref{BN_Res}d), which are consistent with the expected monofractal properties of the original time series within the expected limitations of numerical finite sample estimates.

\subsection{Binomial multifractal cascade}
The binomial multifractal cascade (BMC) is one of the most representative models for which the multifractal features can be derived analytically, allowing a direct comparison with the numerical results of our method (for details on the BMC model, see~\cite{feder2013fractals}). In the following, we explain only briefly the essential details of the model~\cite{feder2013fractals,Peitgen2006,Cheng2014}. 

In order to generate a series $z_j$, with $j=1,\ldots,2^{m_{max}}$, we consider the following expression: 
\begin{equation}
z_j=a^{n(j-1)}(1-a)^{m_{max}-n(j-1)},
\end{equation}
where $a$ $\in (0.5,1)$ is a parameter, $n(j-1)$ is the number of digits equal to 1 in the binary representation of the index $j-1$, and $m_{max}$ represents the number of iterations. One of the advantages of studying binomial cascades is that the key expressions of the multifractal properties can be obtained analytically in a straightforward manner as \cite{MFDFA}
\begin{equation}
\tau(q)=-\frac{\ln[a^q+(1-a)^q]}{\ln(2)},
\label{taubmc}
\end{equation}
\noindent 
and
%\begin{equation}
%		f(\alpha)=q\alpha - \frac{-\ln[a^q+(1-a)^q]}{\ln(2)}. 
%\label{fbmc}
%\end{equation}
\begin{equation}
    \alpha = -\frac{[a^q\ln(a)+(1-a)^q\ln(1-a)]}{\ln(2)[a^q+(1-a)^q]},
\label{fbmc}
\end{equation}
\noindent 
from which the multifractal spectrum can be obtained~\cite{feder2013fractals}. 

\begin{figure}[h]
    \includegraphics[width=0.5\textwidth]{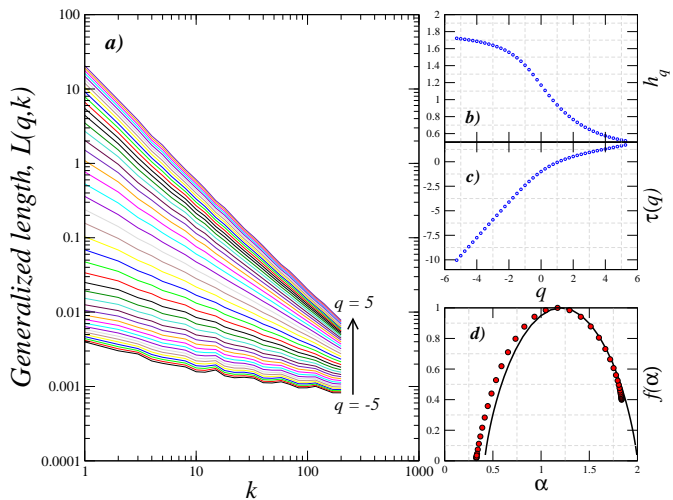}
    \caption{Multifractal Higuchi dimension analysis of a BMC series with $N=2,048$. a) Generalized curve length $\mathcal{L}(k,q)$ as a function of the scale $k$ for $q$-values within the interval $-5 \leq q \leq 5$. b) Scaling exponents $h_q$ as a function of $q$, obtained from the calculations of the slopes ($d_q$) of the curves in a), for scales $1 \leq k \leq 200$. c) Behavior of the mass function $\tau(q)$ as a function of $q$, which indicates a clearly nonlinear dependence. d) Multifractal spectrum $f(\alpha)$ versus $\alpha$. Here, a broad spectrum and a good agreement with the theoretical values (Eq.~\ref{fbmc}) is observed.}    \label{BMC_Res}
\end{figure}

The numerical results of our MF-HDA method for BMC series with parameter $a=0.75$ and $N=2^{m_{max}}=2,048$ are shown in Fig.~\ref{BMC_Res}. The behavior of the generalized Higuchi curve length $\mathcal{L}(k,q)$ versus $k$ for $-5\leq q \leq 5$ is depicted in Fig.~\ref{BMC_Res}a, where evident changes in the slope (i.e., the generalized fractal dimensions $d_q$) are present as $q$ varies. The corresponding $h_q$ exponents are shown in Fig.~\ref{BMC_Res}b. The fact that $h_q$ varies with $q$ evidences the multifractal property of the BMC series. The mass function  $\tau(q)$ (Fig.~\ref{BMC_Res}c) presents a clearly nonlinear behavior with $q$, as expected for multifractal series~\cite{feder2013fractals}. Finally, the resulting multifractal spectrum is shown in Fig.~\ref{BMC_Res}d, which was obtained from $h_q$ and $\tau(q)$ through the Legendre transform. We observe a very good agreement with the analytical predictions of the BMC model, except for an absence of the largest $\alpha$ values which may lead to an underestimation of the multifractal width most likely occurring due to the short length of the considered sequence. %\textcolor{blue}{except for an absence of the largest $\alpha$ values (corresponding to negative moment orders $q$, which are most affected by the employed regularization procedure removing the smallest increments) leading to an underestimation of the total width of the multifractal spectrum}.

\section{Comparison of MF-HDA with MF-DFA}

In the previous section, we have already presented results on the multifractal properties for selected time series with prescribed lengths and moments within a predefined interval of $q$. In the following, we will be interested in comparing our new MF-HDA method with the established MF-DFA approach, which presents one of the most widely recognized and applied methodologies to assess the presence of multifractality in nonstationary time series \cite{Jiang2019,MFDFA,Gieraltowski2012,Reyes2018,Matia2003}. For a detailed study on the applicability of MF-DFA and the WTMM method as another benchmark approach in multifractal analysis, the reader is referred to~\cite{Waveletvsmdfa2006}.  

\subsection{Multifractal detrended fluctuation analysis}

Given a time series $X(t)$, we compute the associated profile \cite{MFDFA} $$Y(t)=\sum_{i=1}^t (X(i)-\bar{X})$$ with $t=1,\ldots,N$, where $N$ is the number of values in the time series and $\bar{X}$ denotes the corresponding mean. Next, the integrated series $Y(t)$ is divided into $N_s$ boxes of equal size $\lfloor N/s\rfloor$. The local trend $Y_r(t)$ is calculated separately for data from each of the segments $r=1,\ldots,N_s$ by least square fitting of a polynomial of prescribed degree (i.e., the order of detrending) and removed from the profile. Then, the mean square fluctuation of the detrended profile in each segment is given as $$F^2(s,r)=\frac{1}{s}\sum_{i=1}^s(Y((r-1)s+i)-Y_{r}(i))^2.$$ The resulting (squared) detrended fluctuation function $F^2(s)$ is defined as the average mean square fluctuation taken over all segments $r$. For monofractal series, its square-root behaves like $F(s)\sim s^H$ with $H$ being an estimate of the process' characteristic Hurst exponent.

To generalize this detrended fluctuation analysis to a multifractal analysis framework, we consider the order-$q$ moments of the mean square fluctuations at scale $s$ by setting
$$F(q,s)=\left(\frac{1}{N_s}\sum_{r=1}^{N_s} [F^2(s,r)]^{q/2}\right)^{1/q}$$ and evaluating this property for several values of $q$ (retaining the standard detrended fluctuation analysis for $q=2$) \cite{MFDFA}. Finally, the scaling behavior is described by $F(q,s)\sim s^{h_q}$, where $h_q$ is the generalized Hurst exponent, which is related to the mass exponent $\tau(q)$ by means of Eq.~\eqref{mass}. 

In the following, we will compare the performance of the resulting MF-DFA method with our new MF-HDA approach by considering (i) the effect of different time series lengths and (ii) different ranges of $q$ values. 

\subsection{Fractional Gaussian noise}

\begin{figure*}[htb]
    \includegraphics[width=\textwidth]{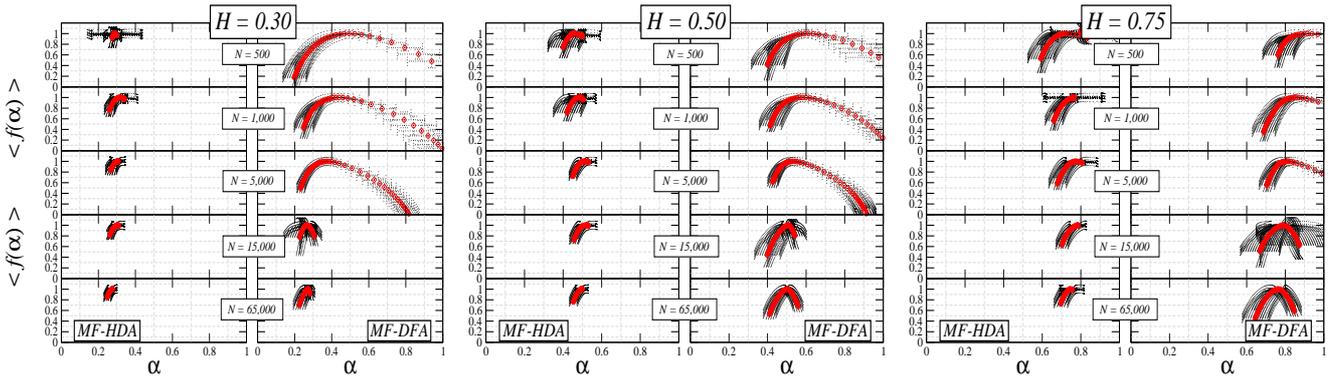}
    \caption{Multifractal spectra estimated by MF-HDA and MF-DFA for fractional Gaussian noise series with different lengths. The cases $H=0.3$ (left), $H=0.5$ (middle) and $H=0.75$ (right) are presented for several representative lengths. The interval for the parameter $q$ was chosen as $-5\leq q\leq 5$. For the MF-HDA estimates, we have removed the percentile $p_r=1$ (see Section.~3 and Appendix B). Symbols and associated error bars indicate the mean and standard deviation from 10 independent realizations.}
    \label{Len_MF_spectrum}
\end{figure*}

As in the previous section, we first consider monofractal time series with prescribed Hurst exponents of $H=0.3$, $H=0.5$ and $H=0.75$ with different lengths $N=500$, $1,000$, $5,000$, $15,000$, and $65,000$. We apply both MF-HDA and MF-DFA to the generated series. Figure~\ref{Len_MF_spectrum} shows the corresponding estimates of the associated multifractal spectra. Especially for short time series lengths, MF-DFA displays wider spectra of values as compared to those obtained by means of MF-HDA. This behavior is especially noticeable for $N\leq 5,000$, where the MF-DFA spectrum is markedly extended towards larger $\alpha$. As the time series length increases, both methods reveal gradually more similar multifractal features, except for the fact that the spectra estimated by means of MF-HDA are less symmetric, due to an absence of the largest $\alpha$ values (corresponding to very negative moment orders $q$), which are most affected by the employed regularization procedure removing the smallest increments. %\textcolor{blue}{which could be corresponding to negative moment orders $q$, which are most affected by the employed regularization procedure removing the smallest increments)could again be a reflection of the effects of the regularization procedure affecting especially the negative moment orders $q$ and, hence, the largest $\alpha$ values}. 

\begin{figure*}[htb]
    \includegraphics[width=\textwidth]{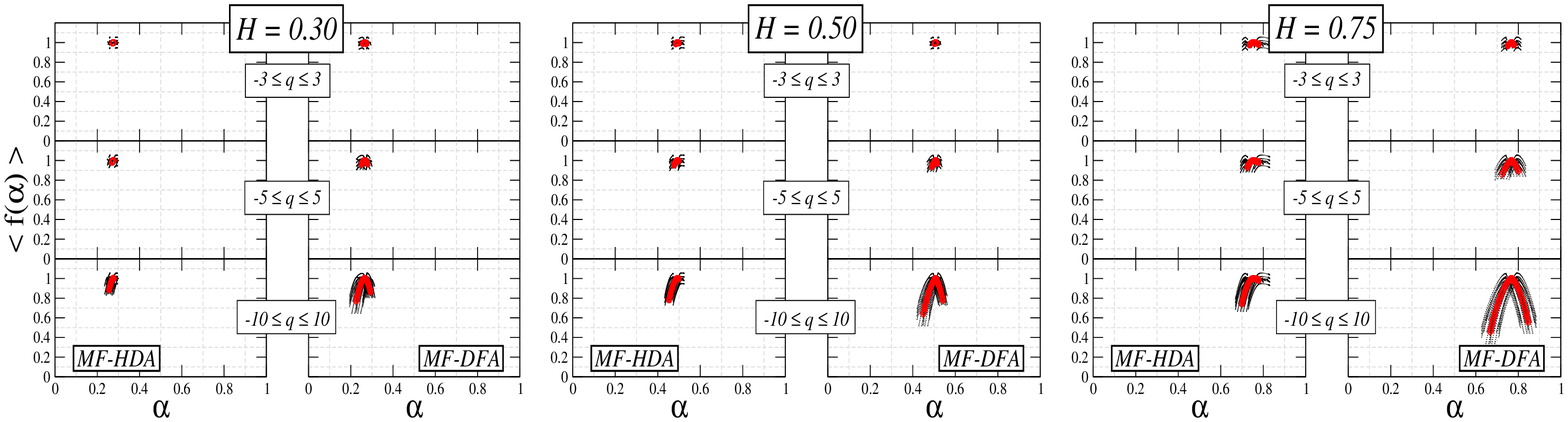}
    \caption{Multifractal spectra estimated by MF-HDA and MF-DFA for fractional Gaussian noise series with different intervals of $q$ values for the three cases $H=0.3$ (left), $H=0.5$ (middle) and $H=0.75$ (right). The results correspond to a fixed time series length of $N=130,000$ and a removal of the percentile $p_r=1$ in MF-HDA. The fitting ranges in the partition function were $k\in[1,N/10]$ and $s\in[200,N/10]$ for MF-HDA and MF-DFA, respectively. Symbols and error bars represent the mean and standard deviation from 10 independent realizations. }
    \label{Q_all}
\end{figure*}

Concerning the effect of the selected range of $q$ values, we also evaluated the changes in  $f(\alpha)$ in terms of different intervals of $q$ for the same monofractal series, here with a fixed length of $N=130,000$. The corresponding results are shown in Fig.~\ref{Q_all}. We observe that, for all three monofractal sequences, both methods lead to similar spectra for the interval $|q|\leq 3$, while for intervals $|q|\leq 5$ and $|q|\leq 10$, MF-HDA provides narrower spectra compared to those obtained by means of MF-DFA for the reasons already mentioned above related to the regularization affecting the smallest fluctuations.

\subsection{Binomial multifractal cascade}

\begin{figure}[htb]
	%\vspace*{1.65cm}
    \includegraphics[width=0.5\textwidth]{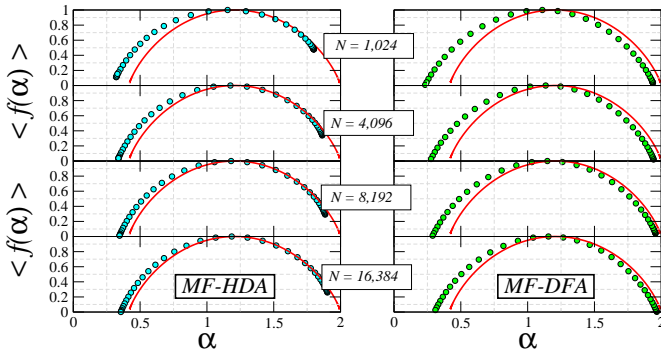}
    \caption{Multifractal spectra estimated by MF-HDA (left) and MF-DFA (right) for the BMC model obtained using different time series lengths. The red lines represent the theoretical values obtained from Eq.~\eqref{fbmc}.}
    \label{Len_BMC}
\end{figure}

In addition to simple monofractal processes, we also compared the MF-HDA and MF-DFA methods for the BMC model for several time series lengths. Figure~\ref{Len_BMC} shows the spectra estimated with both methods along with the analytical results. We find that the obtained estimates are generally in good agreement with the theoretical spectra. Some minor deviations at small $\alpha$ values (i.e., for large positive $q$ values) are observed with both methods (with MF-DFA generally showing slightly larger deviations from the theoretical values), while MF-HDA does not manage to capture the rightmost part of the spectra (large $\alpha$ associated with large negative $q$ values) especially for very short time series lengths.

\begin{figure}[htb]
	%\vspace*{0.985cm}
    \includegraphics[width=0.5\textwidth]{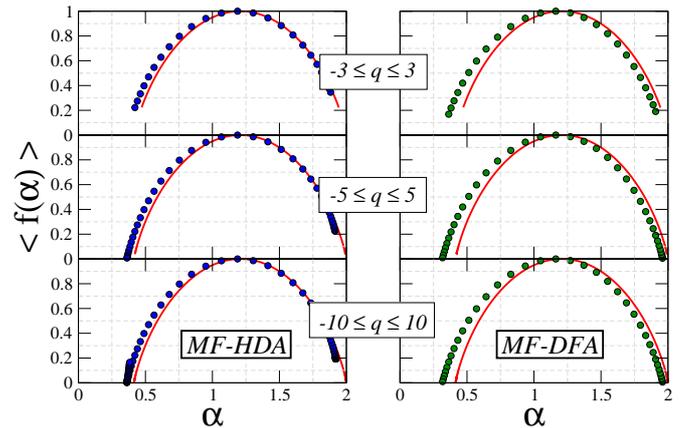}
    \caption{Multifractal spectra estimated by MF-HDA and MF-DFA for the BMC model at the 15-th iteration (length $N=32,768$) using different intervals of $q$ values. The fitting region of the partition function has been selected as $k\in[1,N/10]$.}
    \label{Q_BMC}
\end{figure}

We also tested the dependence of the obtained estimates on the range of $q$ values considered. Figure~\ref{Q_BMC} shows that, as expected, the effect of widening the $q$ interval is to increase the width of the spectrum, with fairly similar behaviors of both methods. 

\begin{figure}[htb]
    \includegraphics[width=0.5\textwidth]{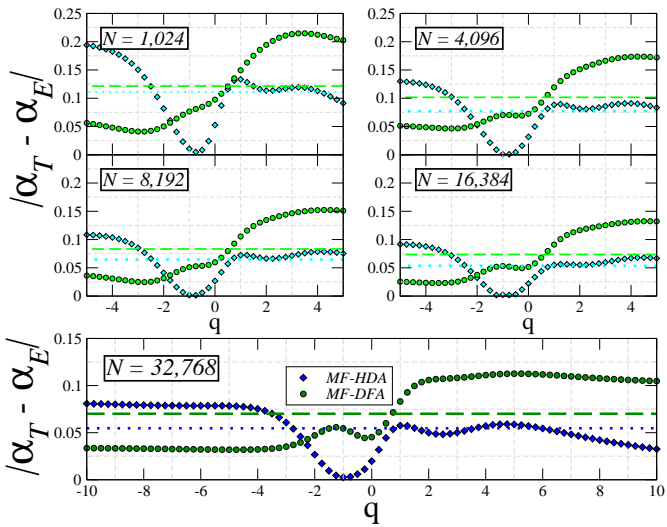}
    \caption{Absolute differences of local scaling exponents between theoretical ($\alpha _ T$) and estimated ($\alpha _ E$) values of the BMC model obtained by either MF-HDA or MF-DFA as functions of the associated moment order $q$ for several representative time series lengths. The circles (blue) and diamonds (green) correspond to the absolute differences from the spectra in Figs.~\ref{Len_BMC} and \ref{Q_BMC}. For $q>-1$, the bias of the scaling exponents is smaller for MF-HDA as compared to MF-DFA, while the opposite applies for $q<-2$, suggesting that the results from MF-HDA are slightly better representing the properties of large fluctuations (positive moments) while MF-DFA performs better for small ones (negative moments). For visual comparison, the average absolute biases over the entire interval are represented for both MF-HDA (dotted lines) and MF-DFA (dashed lines).}
    \label{Delta_a_Err}
\end{figure}

For a more systematic comparison between both approaches, we finally computed the absolute differences between the theoretical values of the singularity strength exponents $\alpha_T$ (obtained from Eqs.~\eqref{taubmc} and \eqref{fbmc}) and the corresponding numerical estimates $\alpha_E$ obtained using either MF-HDA or MF-DFA. Figure~\ref{Delta_a_Err} shows these absolute differences (i.e., the estimation bias) as a function of the moments $q$ for several representative time series lengths. For very short series and for $q\geq -1$, the MF-HDA estimates are closer to the theoretical values, while differences between the theoretical and estimated values are smaller for $q\leq -2$ when using MF-DFA (lacking the shortcomings due to the regularization procedure within our numerical implementation of MF-HDA). For long series, both methods lead to similar results with slightly smaller deviations when using the new MF-HDA method. From these results, we conclude that in the BMC model MF-HDA exhibits certain advantages (lower bias) over MF-DFA especially for short time series and when only a small interval of $q$ values is considered. 

\section{Real-world application: Word length sequences}

As an application to empirical data sets with intrinsic (time) order, we consider the sequences of subsequent word lengths (i.e., the numbers of letters forming each word \cite{guzman2015word}) obtained from a selection of fictional texts in English language, which include mainly novels from different authors (see Tab.~\ref{Tab_books} for details). Previous studies have highlighted the presence of positive correlations and multifractality in such word length sequences, which have been attributed to the fact that written language is the conformation of grammatical properties and semantic connotations with the idea of expressing information \cite{Piantadosi3526}. %\textcolor{blue}{Lev: could you maybe add a specific reference for the last sentence?} DONE

In order to allow for a feasible application of the MF-HDA method to the different word length series, the original length sequence has first been integrated to obtain a profile compatible with a fractional Brownian motion regime. Then, we applied both MF-HDA and MF-DFA to word length sequences with different segment lengths ($N=2,500,\ 5,000,\ 10,000$). In order to improve the reliability of the obtained statistics, several non-overlapping segments of the same length $N$ have been considered for each book. Then, we have used the following procedure to perform MF-HDA: 
\begin{enumerate}[(i)]
\item For each $q$ and the $m$-th segment, the generalized curve length $\mathcal{L}_m(k,q)$ is calculated. 
\item The average value $\mathcal{L}_{av}(k,q)=\frac{1}{M}\sum_{m=1}^{M} \mathcal{L}_m(k,q)$ is evaluated, where $M$ denotes the number of segments for each book. 
\item The value $d_{q,av}$ is estimated from the fit of the scaling behavior $\mathcal{L}_{av}(k,q)\sim k^{d_{q,av}}$.
\end{enumerate}

\begin{figure}
    \includegraphics[width=0.5\textwidth]{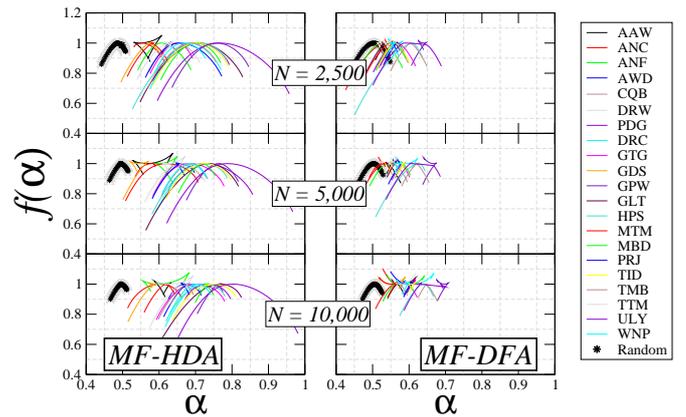}
    \caption{Averaged multifractal spectra $f(\alpha)$ for word length sequences from several books and different segment sizes. (left) Multifractal spectra estimated using MF-HDA. The average generalized curve length from non-overlapping segments was used to estimate the scaling exponent $d_{q,ave}$ with a fitting interval $25\leq k\leq T/10$, where $T$ is the segment length. For estimating the spectra, we have removed the percentile $p_r=15$ (see Section 3 and Appendix B), while qualitatively similar results were obtained for values $p_r=5$ and $p_r=10$ (not shown). The results indicate a variety of levels of multifractality. The range of scaling exponents ($0.5\leq \alpha < 1$) indicates that the accumulated word length series exhibit correlated behavior, while for shuffled and integrated increments very narrow spectra centred at $h=0.5$ are observed. (Right) Multifractal spectra estimated using MF-DFA. A similar procedure as for MF-HDA was used to estimate the scaling exponents $h_{q,av}$ \cite{Matia2003}. The fitting region was $50\leq s \leq T/10$.   }
    \label{Word_len}
\end{figure}

Figure~\ref{Word_len} shows the results obtained with both MF-HDA and MF-DFA. We observe that unlike for the simple stochastic model cases studied in the previous section, the spectra $f(\alpha)$ obtained with MF-HDA are broader than those provided by MF-DFA, especially for short sequences. Notoriously, the expected convex shape of multifractal spectra is not well expressed in the MF-DFA results. Moreover, as the length of the considered segments is increased, the width of the spectra decreases but still remains relatively broad for MF-HDA as compared to MF-DFA. This result is remarkable since, as we have noticed in our controlled numerical experiments above, we may have expected MF-DFA tending to rather overestimate the multifractal spectral width for short time series. 

As a consistency check, we finally generated shuffled versions of the word length series by applying a random permutation of the individual word lengths, thereby destroying eventual correlations in the sequence. The results of applying both multifractal analysis methods to the accordingly randomized sequences are also presented in Fig.~\ref{Word_len}. As expected, we obtain narrow spectra centered at $\alpha=0.5$ for all books and different segment lengths, which do not differ much between the two methods.

\begin{table*}
	\centering
	\caption{Books used for the multifractal analysis of word length sequences. The table lists  the title and author, code name and the total number of words, along with estimates of the global Hurst exponent $h_{q=1}$, the spectral mode $\alpha^*$ and the spectral width $ \Delta \alpha$ as obtained using MF-HDA averaged over different segments with length $N=2,500$ (see Fig.~\ref{Word_len}). Although we have not followed a particular selection strategy for the studied titles beyond just considering well-known books, all estimated values for the global Hurst exponent ($h_{q=1}$) are clearly larger than 0.5, confirming the presence of long range correlations in the word length series. These values are also in good agreement with previous monofractal analyses reported in~\cite{guzman2015word,Montemurro2002,Rodriguez2014}. The $\alpha^*$ values at which the multifractal spectra take their maxima are also consistently larger than $0.5$ for all books. Moreover, most of the books are characterized by relatively broad multifractal spectra with $\Delta \alpha \geq 0.1 $, except for AAW, ANC and TTM (highligthed in bold face), for which the spectra exhibit a discontinuous behavior. Our results are consistent with previous analyses based on traditional multifractal procedures~\cite{Ausloos2012,Chatzigeorgiou2017} and provide a better characterization of the complexity displayed in written texts. Interestingly, the book with the largest multifractal spectral width ($\Delta \alpha=0.36$) corresponds to Joyce's Ulysses (ULY, underlined), which has been described as a text with particularly great diversity in language~\cite{Rice1994}.
  }
    \label{Tab_books}
    \begin{tabular}{|l|c|r|c|c|c|c|}
        \hline
        \textbf{Title and Author} & \textbf{Code} & \textbf{Words} & $h_{q=1}$ & \textbf{$\alpha^*$} & \textbf{$\Delta \alpha$} \\
        \hline
        Alice's Adventure in Wonderland, L. Caroll  & AAW & 27,330 & 0.56 & 0.61 & \textbf{0.08}  \\
        The Anarchy, W. Dalrymple & ANC & 15,971 & 0.54 & 0.55 & \textbf{0.08} \\
        Animal Farm, G. Orwell & ANF & 30,384 & 0.60 & 0.61 & 0.15 \\
        Around the World in 80 Days, J. Verne & AWD & 63,760 & 0.62 & 0.63 & \textbf{0.04} \\
        %The Art of War, Sun Tzu & ART & 10,990 & 0.60 & 0.63 & 0.15 \\
        The Conquest of Bread, P. Kropotkin & CQB & 72,017 & 0.63 & 0.65 & 0.12 \\
        On the Origin of Species, C. Darwin & DRW & 156,812 & 0.62 & 0.64 & 0.17 \\
        The Picture of Dorian Gray, O. Wilde & PDG & 80,408 & 0.68 & 0.70 & 0.21 \\
        Dracula, B. Stoker & DRC & 162,317 & 0.67 & 0.68 & 0.20 \\
        The Great Gatsby, F. S. Fitzgerald & GTG & 50,103 & 0.64 & 0.65 & 0.12 \\
        Golden State, Ben H. Winters & GDS & 27,731 & 0.57 & 0.67 & 0.11 \\
        The Grapes of Wrath, J. Steinbeck & GPW & 187,579 & 0.74 & 0.76 & 0.21 \\
        Gulliver’s Travels, J. Swift & GLT & 104,798 & 0.69 & 0.71 & 0.28 \\
        Hopscotch, J. Cortázar & HPS & 195,703 & 0.69 & 0.71 & 0.26 \\
        %Mein Kampf, A. Hitler, & MNK & 273,388 & 0.60 & 0.61 & 0.11 \\
        %The Communist Manifesto, K. Marx and F. Engels & MNF & 11,537 \\
        The Metamorphosis, F. Kafka & MTM & 22,383 & 0.58 & 0.58 & 0.11 \\
        Moby Dick, H. Melville & MBD & 218,705 & 0.69 & 0.70 & 0.16 \\
        Pierre and Jean, G. de Maupassant & PRJ & 46,544 & 0.64 & 0.66 & 0.21 \\
        The Idiot, F. Dostoyevsky & TID & 247,953 & 0.69 & 0.70 & 0.21 \\
        Three Men in a Boat, J. K. Jerome & TMB & 68,805 & 0.67 & 0.69 & 0.19 \\
        The Time Machine, H. G. Wells & TTM & 32,776 & 0.63 & 0.64 & \textbf{0.05} \\
        Ulysses, J. Joyce & ULY & 272,416 & 0.74 & 0.77 & \underline{0.36} \\
        War and Peace, L. Tolstoy & WNP & 572,628 & 0.66 & 0.67 & 0.14 \\
        War of the Worlds, H. G. Wells & WOW & 60,897 & 0.68 & 0.70 & 0.22 \\
        \hline
    \end{tabular}
\end{table*}

\section{Discussion}

The identification of multifractal properties in time series is a challenge especially when dealing with short sequences, as are present in many real-world cases. The proposed MF-HDA method offers an alternative to assess the multifractal properties of relatively short series by constructing generalized curve lengths, the scaling of which encodes the multifractal characteristics of the signal. The implementation of our methodology is relatively simple, as it is based on the expected $q$-moment values of absolute increment distributions. 

We have described several cases where the MF-HDA method identifies a multifractal scaling behavior of the time series with very similar results as the celebrated MF-DFA approach, which provides a well-established methodology for the analysis of non-stationary signals.  For very short series, the results for the MF-HDA indicate a certain gain in accuracy as compared to those obtained using the MF-DFA method, except for very negative moment orders $q$.

As with the majority of studies focusing on the detection of multifractality in non-stationary time series, the proposed methodology is subject to certain limitations: (i) dominant scaling exponents of the time series can be located outside of the region where the MF-HDA allows us to detect multifractal scaling in a reliable and robust manner; (ii) the limited sampling rate and numerical precision of the time series makes the expected moments of the generalized curve length quite difficult to estimate (especially for negative moments); (iii) the removal of the $p_r$ percentile of the absolute increments has a regularizing effect but can potentially lead to another bias in the estimated scaling exponents, especially for very negative $q$ values. Further refining the presented numerical procedures to address the aforementioned challenges should be addressed in follow-up studies.

 \section{Conclusions}
 We have presented a generalization of the Higuchi fractal dimension analysis to a multifractal framework. The main advantage of the new MF-HDA method in comparison with other established multifractal analysi techniques lies in the property of obtaining stable estimates of multifractal scaling characteristics even when the analyzed sequences are relatively short. However, certain precautions must be considered when analyzing real-world time series, for which the origin of multifractality can be associated with different factors such as power law correlations or a heavy-tailed probability density function of the data \cite{Kantelhardt2011,MFDFA}. For a more reliable estimation of multifractal characteristics, an integrated approach involving complementary methods such as MF-HDA, MF-DFA or WTMM is recommended. We envision to further explore the corresponding potentials and more systematically apply our methodology to a range of real-world and simulated time series in our future work.  
 
 \section*{Code availablity}
 Numerical implementations and examples for the application of MF-HDA can be found at \url{https://github.com/carrizales90/MF-HDA}.
 %\textit{GitHub} \cite{carrizales2021}.  \todo[inline]{Do we need to mention why we do not consider a symmetric truncation (i.e., a two-sided trimmed mean)?}

\begin{acknowledgements}
This  work  was  partially  supported  by  programs  EDI  and  COFAA  from  Instituto Polit\'ecnico Nacional and Consejo Nacional de Ciencia y Tecnolog\'ia, M\'exico. 
RVD has been partially supported by the Federal Ministry for Education and Research of Germany (BMBF) via the JPI Climate/JPI Oceans project ROADMAP (grant no. 01LP2002B). 
We thank F. Angulo-Brown, D. Aguilar-Velazquez, I. Reyes-Ram\'irez and C. Reyes-Manzano for useful discussions and suggestions.
\end{acknowledgements}


%\section*{Appendix}
\appendix
\section{The case $q = 0$}
In multifractal analysis, it is common that the scaling exponent $d_q$ is not well defined when $q \rightarrow{} 0$. In our case, this value cannot be directly determined by means of the generalized curve lengths (Eq.~\eqref{Lkq}) due to the presence of a divergence in the exponent. More formally, we have

\begin{equation}
\begin{split}
    \lim_{q\to0}  \mathcal{L}(q, k) = \lim_{q\to0} \frac{N-1}{k^2} \Bigg\{ \sum_{n=1}^{N_b} \langle (\Delta X_n(k))^q \rangle P_n(\Delta X(k)) \Bigg\}^{1/q} 
    \\ \sim \lim_{q\to0} k ^ {- d_q}.
\end{split}
\end{equation}
\noindent
Using some algebraic operations applied to the latter equation, which are omitted here for brevity, and applying L'H$\hat{o}$pital's rule, we find that a logarithmic transformation is required in order to determine the scaling exponent $d_0$ as
\begin{equation}
    \mathcal{L}(0, k) \equiv \frac{N-1}{k^2} \exp{ \Bigg( E [ \langle \ln \{ \Delta X(k) \} \rangle ] \Bigg) } \sim k^{-d_0},
\end{equation}
\noindent
where $E [ \langle \ln \{ \Delta X(k) \}  \rangle ]= \sum_{n=1}^{N_b} \langle \ln \{ \Delta X(k)\} \rangle P_n(\Delta X(k))$.

\section{Regularization effect of removing the lower percentile of absolute increments}

As discussed in Section~\ref{Sec-MDA}, numerical instabilities can appear in the evaluation of the moments of the generalized curve length $\mathcal{L}(k,q)$ (Eq.~\ref{Lkq}), especially for $q<-1$. In this case, we have suggested that the local mean could be replaced by a one-sided trimmed version $\langle \Delta X_{n'}(k)\rangle _{>p_r}$ in Eq.~\eqref{Lkq}, where $n'$ represents the $n'$--th interval of a new equiprobable partition in which we have removed the $r$--th percentile $p_r$ of the empirical distribution of the absolute increments. We note that numerical experiments with both, one-sided (asymmetrically) and two-sided (symmetrically) trimmed means revealed no qualitative differences in the resulting estimates (not shown), while removing the uppermost percentiles (i.e., very large increments) appears unnecessary since those values have no negative effects on the stability of the numerical estimates of the generalized curve lengths.

To address the problem of selecting a specific percentile to be removed, we focus here on just one statistical property, the confidence interval (CI) of the estimated slope (i.e., the scaling exponent $d_q$ in Eq.~\eqref{Scal}) in the linear regression of the double-logarithmic generalized curve length versus scale relationship, at a certain confidence level (here, $\alpha=0.05$), and for the most negative value of $q$. For two-sided confidence intervals, the CI width (CIW) (measured in units of the associated standard error) is given by $\hat{I}_{q,p_{r}} \equiv I_{q,p_{r}}/S_{d_q}$, with $S_{d_{q}}$ being the standard error of the estimated slope $d_q$ \cite{wasserman2013all}. 

\begin{figure}
	\vspace*{0.5cm}
    \includegraphics[width=0.5\textwidth]{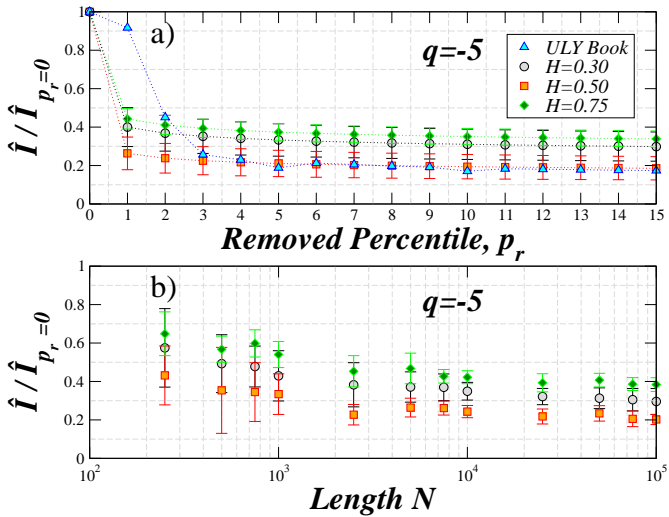}
    \caption{(a) Behavior of the rescaled confidence interval in dependence on the removed percentil $p_r$ of the distribution of absolute increments for $q=-5$ for fractional Gaussian noises and word length data (ULY book). (b) Dependence of the rescaled confidence interval on the sequence length $N$ for $q=-5$ in the case of fractional Gaussian noises. As $N$ increases, $\hat{I}/\hat{I}_{p_r=0}$ decreases and becomes independent of the system size. Error bars indicate the standard deviation estimated from 10 independent realizations.}
    \label{BN_W-LN_Perc}
\end{figure}

Figure~\ref{BN_W-LN_Perc} shows the behavior of the rescaled CIW (in units of $\hat{I}_{p_{r}=0}$) as a function of the removed percentile $p_r$, for some of the simulated stochastic processes and real-world data sets discussed in Sections~4 and 6, respectively, for $q=-5$. The results show that, as the removed percentile is increased, the CIW decreases in such a way that, for fractional Gaussian noises with $H=0.3$, $H=0.5$ and $H=0.75$, the rescaled CIW has decayed by more than one half of its initial value when $p_r=1$, while for the world length data (exemplified here by the ULY book) the observed decay is slower. For practical purposes, we suggest that a criterion for selecting the value of the percentile to be removed should consider empirically a value of the percentile for which the rescaled CIW has stabilized, that is, even if higher percentiles are removed, there are no substantial further changes. We observe that $p_r\approx 5$ and $p_r \approx 12$ would be desirable in the cases of the simulated fractional Gaussian noises and word length data, respectively. 

While the suggested strategy presents just a first attempt to improving the practical estimation of the generalized Higuchi fractal dimensions, we emphasize that there may be cases in which the rescaled CIW may behave in a more unstable way with increasing percentile $p_r$. In such cases, additional numerical tests with larger $p_r$ values may become necessary to determine a reliable value leading to sufficiently stable estimates.
%\textbf{We also warn that in some cases, the behavior of the rescaled CIL may be unstable with growth, and additional tests with higher values in $p_r$ are necessary to determine the stability value.}

\bibliographystyle{unsrt}
\bibliography{biblio}   % name your BibTeX data base

\begin{thebibliography}{10}

\bibitem{mandelbrot1983fractal}
B.B. Mandelbrot.
\newblock {\em The fractal geometry of nature}, volume~1.
\newblock W.H. Freeman New York, 1982.

\bibitem{janssen1998statistics}
M.~Janssen.
\newblock Statistics and scaling in disordered mesoscopic electron systems.
\newblock {\em Physics Reports}, 295(1-2):1--91, 1998.

\bibitem{feder2013fractals}
J.~Feder.
\newblock {\em Fractals}.
\newblock Springer Science \& Business Media, 2013.

\bibitem{frisch1985singularity}
U.~Frisch and G.~Parisi.
\newblock On the singularity structure of fully developed turbulence.
\newblock In M.~Ghil, R.~Benzi, and G.~Parisi, editors, {\em {Turbulence and
  Predictability in Geophysical Fluid Dynamics and Climate Dynamics}}, pages
  84--88. North Holland, New York, 1985.

\bibitem{Stanley1988}
H.E. Stanley and P.~Meakin.
\newblock Multifractal phenomena in physics and chemistry.
\newblock {\em Nature}, 335(6189):405--409, 1988.

\bibitem{Ivanov1999}
P.Ch. Ivanov, L.A.~N. Amaral, A.L. Goldberger, S.~Havlin, M.G. Rosenblum, Z.R.
  Struzik, and H.E. Stanley.
\newblock Multifractality in human heartbeat dynamics.
\newblock {\em Nature}, 399(6735):461--465, 1999.

\bibitem{Schmitt2016}
F.G. Schmitt and Y.~Huang.
\newblock {\em {Stochastic Analysis of Scaling Time Series}}.
\newblock Cambridge University Press, Cambridge, 2016.

\bibitem{Jiang2019}
Z.Q. Jiang, W.J. Xie, W.X. Zhou, and D.~Sornette.
\newblock Multifractal analysis of financial markets: a review.
\newblock {\em Reports on Progress in Physics}, 82(12):125901, 2019.

\bibitem{Kantelhardt2011}
J.W. Kantelhardt.
\newblock {\em Fractal and Multifractal Time Series}, pages 463--487.
\newblock Springer New York, New York, NY, 2011.

\bibitem{bunde2012science}
A.~Bunde, J.~Kropp, and H.J. Schellnhuber.
\newblock {\em The science of disasters: climate disruptions, heart attacks,
  and market crashes}.
\newblock Springer Science \& Business Media, 2012.

\bibitem{kolmogorov1941local}
A.N. Kolmogorov.
\newblock The local structure of turbulence in incompressible viscous fluid for
  very large reynolds numbers.
\newblock {\em Cr Acad. Sci. URSS}, 30:301--305, 1941.

\bibitem{frisch1995turbulence}
U.~Frisch and A.N. Kolmogorov.
\newblock {\em Turbulence: the legacy of AN Kolmogorov}.
\newblock Cambridge University Press, Cambridge, 1995.

\bibitem{mccauley1990introduction}
J.L. McCauley.
\newblock Introduction to multifractals in dynamical systems theory and fully
  developed fluid turbulence.
\newblock {\em Physics Reports}, 189(5):225--266, 1990.

\bibitem{coleman1992introduction}
P.H. Coleman and L.~Pietronero.
\newblock Introduction to multifractals in dynamical systems theory and fully
  developed fluid turbulence.
\newblock {\em Physics Reports}, 213(6):311--389, 1992.

\bibitem{mandelbrot1989multifractal}
B.B. Mandelbrot.
\newblock Multifractal measures, especially for the geophysicist.
\newblock In {\em Fractals in Geophysics}, pages 5--42. Springer, 1989.

\bibitem{mandelbrot1999multifractal}
B.B. Mandelbrot.
\newblock A multifractal walk down wall street.
\newblock {\em Scientific American}, 280(2):70--73, 1999.

\bibitem{paladin1987anomalous}
G.~Paladin and A.~Vulpiani.
\newblock Anomalous scaling laws in multifractal objects.
\newblock {\em Physics Reports}, 156(4):147--225, 1987.

\bibitem{olemskoi2000theory}
A.I. Olemskoi and V.F. Klepikov.
\newblock The theory of spatiotemporal pattern in nonequilibrium systems.
\newblock {\em Physics Reports}, 338(6):571--677, 2000.

\bibitem{touchette2009large}
H.~Touchette.
\newblock The large deviation approach to statistical mechanics.
\newblock {\em Physics Reports}, 478(1-3):1--69, 2009.

\bibitem{kwapien2012physical}
J.~Kwapie{\'n} and S.~Dro{\.z}d{\.z}.
\newblock Physical approach to complex systems.
\newblock {\em Physics Reports}, 515(3-4):115--226, 2012.

\bibitem{Anselmet1984}
F.~Anselmet, Y.~Gagne, E.J. Hopfinger, and R.A. Antonia.
\newblock High-order velocity structure functions in turbulent shear flows.
\newblock {\em Journal of Fluid Mechanics}, 140:63--89, 1984.

\bibitem{barabasi1991multifractality}
A.L. Barab{\'a}si and T.~Vicsek.
\newblock Multifractality of self-affine fractals.
\newblock {\em Physical Review A}, 44(4):2730, 1991.

\bibitem{MMWT}
J.~F. Muzy, E.~Bacry, and A.~Arneodo.
\newblock Wavelets and multifractal formalism for singular signals: Application
  to turbulence data.
\newblock {\em Physical Review Letters}, 67:3515--3518, 1991.

\bibitem{muzy1994multifractal}
J.F. Muzy, E.~Bacry, and A.~Arneodo.
\newblock The multifractal formalism revisited with wavelets.
\newblock {\em International Journal of Bifurcation and Chaos}, 4(02):245--302,
  1994.

\bibitem{Welter2013}
G.S. Welter and P.A.A. Esquef.
\newblock Multifractal analysis based on amplitude extrema of intrinsic mode
  functions.
\newblock {\em Physical Review E}, 87:032916, 2013.

\bibitem{Alberti2019}
T.~Alberti, G.~Consolini, V.~Carbone, E.~Yordanova, M.F. Marcucci, and
  P.~De~Michelis.
\newblock Multifractal and chaotic properties of solar wind at mhd and kinetic
  domains: An empirical mode decomposition approach.
\newblock {\em Entropy}, 21(3):320, 2019.

\bibitem{MFDFA}
J.~W. Kantelhardt, S.A. Zschiegner, E.~Koscielny-Bunde, S.~Havlin, A.~Bunde,
  and H.E. Stanley.
\newblock Multifractal detrended fluctuation analysis of nonstationary time
  series.
\newblock {\em Physica A: Statistical Mechanics and its Applications},
  316(1–4):87 -- 114, 2002.

\bibitem{Gieraltowski2012}
J.~Giera{\l}towski, J.J. {\.Z}ebrowski, and R.~Baranowski.
\newblock Multiscale multifractal analysis of heart rate variability recordings
  with a large number of occurrences of arrhythmia.
\newblock {\em Physical Review E}, 85(2):021915, 2012.

\bibitem{Waveletvsmdfa2006}
P.~O\ifmmode \acute{s}\else \'{s}\fi{}wi\ifmmode~\mbox{\c{e}}\else
  \c{e}\fi{}cimka, J.~Kwapie\ifmmode~\acute{n}\else \'{n}\fi{}, and
  S.~Dro\ifmmode \dot{z}\else \.{z}\fi{}d\ifmmode~\dot{z}\else \.{z}\fi{}.
\newblock Wavelet versus detrended fluctuation analysis of multifractal
  structures.
\newblock {\em Physical Review E}, 74:016103, Jul 2006.

\bibitem{higuchi1988approach}
T.~Higuchi.
\newblock Approach to an irregular time series on the basis of the fractal
  theory.
\newblock {\em Physica D: Nonlinear Phenomena}, 31(2):277 -- 283, 1988.

\bibitem{Higuchi1990}
T.~Higuchi.
\newblock Relationship between the fractal dimension and the power law index
  for a time series: a numerical investigation.
\newblock {\em Physica D: Nonlinear Phenomena}, 46(2):254--264, 1990.

\bibitem{Nikolopoulos2020}
D.~Nikolopoulos, E.~Petraki, P.H. Yannakopoulos, G.~Priniotakis, I.~Voyiatzis,
  and D.~Cantzos.
\newblock {Long-Lasting Patterns in 3 kHz Electromagnetic Time Series after the
  ML= 6.6 Earthquake of 2018-10-25 near Zakynthos, Greece}.
\newblock {\em Geosciences}, 10(6):235, 2020.

\bibitem{Ramirez2008}
A.~Ram{\'\i}rez-Rojas, E.L. Flores-M{\'a}rquez, L.~Guzman-Vargas,
  G.~G{\'a}lvez-Coyt, L.~Telesca, and F.~Angulo-Brown.
\newblock {Statistical features of seismoelectric signals prior to M7.4
  Guerrero-Oaxaca earthquake (M{\'e}xico)}.
\newblock {\em Natural Hazards and Earth System Sciences}, 8(5):1001--1007,
  2008.

\bibitem{Donner2015}
R.V. Donner, S.M. Potirakis, S.M. Barbosa, J.A.O. Matos, A.J.S.C. Pereira, and
  L.J.P.F. Neves.
\newblock Intrinsic vs. spurious long-range memory in high-frequency records of
  environmental radioactivity.
\newblock {\em The European Physical Journal Special Topics}, 224(4):741--762,
  2015.

\bibitem{Cuomo1999}
V.~Cuomo, V.~Lapenna, M.~Macchiato, C.~Serio, and L.~Telesca.
\newblock {Stochastic behaviour and scaling laws in geoelectrical signals
  measured in a seismic area of southern Italy}.
\newblock {\em Geophysical Journal International}, 139(3):889--894, 1999.

\bibitem{Kesic2016}
S.~Kesi{\'c} and S.Z. Spasi{\'c}.
\newblock {Application of Higuchi's fractal dimension from basic to clinical
  neurophysiology: a review}.
\newblock {\em Computer Methods and Programs in Biomedicine}, 133:55--70, 2016.

\bibitem{Guzman2003}
L.~Guzman-Vargas and F.~Angulo-Brown.
\newblock Simple model of the aging effect in heart interbeat time series.
\newblock {\em Physical Review E}, 67(5):052901, 2003.

\bibitem{Schmitt2007}
D.T. Schmitt and P.Ch. Ivanov.
\newblock Fractal scale-invariant and nonlinear properties of cardiac dynamics
  remain stable with advanced age: a new mechanistic picture of cardiac control
  in healthy elderly.
\newblock {\em American Journal of Physiology-Regulatory, Integrative and
  Comparative Physiology}, 293(5):R1923--R1937, 2007.

\bibitem{Contreras2017}
T.J. Contreras-Uribe, L.I. Garay-Jim{\'e}nez, and L.~Guzm{\'a}n-Vargas.
\newblock A point process analysis of electrogastric variability.
\newblock {\em Chaos, Solitons \& Fractals}, 94:16--22, 2017.

\bibitem{graham1989concrete}
R.L. Graham, D.E. Knuth, O.~Patashnik, and S.~Liu.
\newblock Concrete mathematics: a foundation for computer science.
\newblock {\em Computers in Physics}, 3(5):106--107, 1989.

\bibitem{mandelbrot2002gaussian}
B.~Mandelbrot.
\newblock {\em {Gaussian Self-Affinity and Fractals: Globality, the Earth, 1/f
  noise, and R/S}}, volume~8.
\newblock Springer Science \& Business Media, 2002.

\bibitem{Guzman2005}
L.~Guzman-Vargas, A.~Munoz-Diosdado, and F.~Angulo-Brown.
\newblock Influence of the loss of time-constants repertoire in pathologic
  heartbeat dynamics.
\newblock {\em Physica A: Statistical Mechanics and its Applications},
  348:304--316, 2005.

\bibitem{rangarajan2000integrated}
G.~Rangarajan and M.~Ding.
\newblock Integrated approach to the assessment of long range correlation in
  time series data.
\newblock {\em Physical Review E}, 61(5):4991, 2000.

\bibitem{Witt2013}
A.~Witt and B.D. Malamud.
\newblock Quantification of long-range persistence in geophysical time series:
  Conventional and benchmark-based improvement techniques.
\newblock {\em Surveys in Geophysics}, 34(5):541--651, 2013.

\bibitem{Peitgen2006}
H.O. Peitgen, H.~J{\"u}rgens, and D.~Saupe.
\newblock {\em Chaos and fractals: new frontiers of science}.
\newblock Springer Science \&amp; Business Media, 2006.

\bibitem{Cheng2014}
Q.~Cheng.
\newblock Generalized binomial multiplicative cascade processes and
  asymmetrical multifractal distributions.
\newblock {\em Nonlinear Processes in Geophysics}, 21(2):477--487, 2014.

\bibitem{Reyes2018}
C.F. Reyes-Manzano, C.~Lerma, J.C. Echeverr{\'\i}a, M.~Mart{\'\i}nez-Lavin,
  L.A. Mart{\'\i}nez-Mart{\'\i}nez, O.~Infante, and L.~Guzm{\'a}n-Vargas.
\newblock Multifractal analysis reveals decreased non-linearity and stronger
  anticorrelations in heart period fluctuations of fibromyalgia patients.
\newblock {\em Frontiers in physiology}, 9:1118, 2018.

\bibitem{Matia2003}
K.~Matia, Y.~Ashkenazy, and H.E. Stanley.
\newblock Multifractal properties of price fluctuations of stocks and
  commodities.
\newblock {\em EPL (Europhysics Letters)}, 61(3):422, 2003.

\bibitem{guzman2015word}
L.~Guzm{\'a}n-Vargas, B.~Obreg{\'o}n-Quintana, D.~Aguilar-Vel{\'a}zquez,
  R.~Hern{\'a}ndez-P{\'e}rez, and L.S. Liebovitch.
\newblock Word-length correlations and memory in large texts: A visibility
  network analysis.
\newblock {\em Entropy}, 17(11):7798--7810, 2015.

\bibitem{Piantadosi3526}
S.T. Piantadosi, H.~Tily, and E.~Gibson.
\newblock Word lengths are optimized for efficient communication.
\newblock {\em Proceedings of the National Academy of Sciences},
  108(9):3526--3529, 2011.

\bibitem{Montemurro2002}
M.A. Montemurro and P.A. Pury.
\newblock Long-range fractal correlations in literary corpora.
\newblock {\em Fractals}, 10(04):451--461, 2002.

\bibitem{Rodriguez2014}
E.~Rodriguez, M.~Aguilar-Cornejo, R.~Femat, and J.~Alvarez-Ramirez.
\newblock Scale and time dependence of serial correlations in word-length time
  series of written texts.
\newblock {\em Physica A: Statistical Mechanics and its Applications},
  414:378--386, 2014.

\bibitem{Ausloos2012}
M.~Ausloos.
\newblock Generalized hurst exponent and multifractal function of original and
  translated texts mapped into frequency and length time series.
\newblock {\em Physical Review E}, 86(3):031108, 2012.

\bibitem{Chatzigeorgiou2017}
M.~Chatzigeorgiou, V.~Constantoudis, F.~Diakonos, K.~Karamanos,
  C.~Papadimitriou, M.~Kalimeri, and H.~Papageorgiou.
\newblock Multifractal correlations in natural language written texts: Effects
  of language family and long word statistics.
\newblock {\em Physica A: Statistical Mechanics and its Applications},
  469:173--182, 2017.

\bibitem{Rice1994}
T.J. Rice.
\newblock {"Ulysses", Chaos, and Complexity}.
\newblock {\em James Joyce Quarterly}, 31(2):41--54, 1994.

\bibitem{wasserman2013all}
L.~Wasserman.
\newblock {\em All of statistics: a concise course in statistical inference}.
\newblock Springer Science \&; Business Media, 2013.

\end{thebibliography}

%\begin{thebibliography}{}

%\end{thebibliography}

\end{document}